\documentclass[conference]{IEEEtran}
\IEEEoverridecommandlockouts
\usepackage{cite}
\usepackage{amsmath,amssymb,amsfonts}
\usepackage{graphicx}
\usepackage{textcomp}
\usepackage{xcolor}

\usepackage{booktabs}  
\usepackage{amssymb}  
\usepackage{graphicx}  
\usepackage{float}
\usepackage{array}
\usepackage{verbatim}
\usepackage{fancyvrb}
\usepackage{algorithm}
\usepackage{algpseudocode}
\usepackage{amsmath}
\usepackage[position=bottom]{subfig} 
\usepackage{caption}
\usepackage{subcaption}
\usepackage{hyperref}

\def\BibTeX{{\rm B\kern-.05em{\sc i\kern-.025em b}\kern-.08em
    T\kern-.1667em\lower.7ex\hbox{E}\kern-.125emX}}
\begin{document}

\title{FedBGS: A Blockchain Approach to Segment Gossip Learning in Decentralized Systems}

\author{
\IEEEauthorblockN{Fabio Turazza, Marcello Pietri, Marco Picone, Marco Mamei}
\IEEEauthorblockA{\textit{Department of Sciences and Methods for Engineering} \\
\textit{University of Modena and Reggio Emilia, Reggio Emilia, Italy}\\
{name.surname}@unimore.it}
% \and
% %\IEEEauthorblockN{2\textsuperscript{nd} Given Name Surname}
% %\IEEEauthorblockA{\textit{dept. name of organization (of Aff.)} \\
% %\textit{name of organization (of Aff.)}\\
% %City, Country \\
% %email address or ORCID}
% \and
% \and
% \and
}

\maketitle

\begin{abstract}
Privacy-Preserving Federated Learning (PPFL) is a Decentralized machine learning paradigm that enables multiple participants to collaboratively train a global model without sharing their data with the integration of cryptographic and privacy-based techniques to enhance the security of the global system. This privacy-oriented approach makes PPFL a highly suitable solution for training shared models in sectors where data privacy is a critical concern. In traditional FL, local models are trained on edge devices, and only model updates are shared with a central server, which aggregates them to improve the global model. However, despite the presence of the aforementioned privacy techniques, in the classical Federated structure, the issue of the server as a single-point-of-failure remains, leading to limitations both in terms of security and scalability.
This paper introduces FedBGS, a fully Decentralized Blockchain-based framework that leverages Segmented Gossip Learning through Federated Analytics. The proposed system aims to optimize blockchain usage while providing comprehensive protection against all types of attacks, ensuring both privacy, security and non-IID data handling in Federated environments.
\end{abstract}


\section{Introduction}
Over the past decade, owing to the rapid proliferation of digital applications coupled with more affordable and easily implementable data storage solutions, data have acquired substantial legal and economic value. Legally, much of this information is personal and is deemed an inviolable asset for each individual, protected by measures such as the EU GDPR (General Data Protection Regulation) and US HIPAA (Health Insurance Portability and Accountability Act) for sensitive medical information. Economically, companies regard these data as a key asset and are often reluctant to share them. This reluctance creates a sort of “prisoner” dilemma, where data exchange could offer a collective benefit by improving training for a shared model, improving its generalization and reducing biases.

Federated learning was introduced by Google in 2016 through the FedAvg (Federated Averaging) algorithm \cite{mcmahan2023communicationefficientlearningdeepnetworks}, designed to address these issues by establishing a decentralized training process in which data remain with its original owner. In the FedAvg approach, every participant trains a local model and sends only the model updates to a central server. This server aggregates the updates averaging them, to produce a global model, which is then redistributed to the participants. This cycle is repeated over a predefinite number of rounds until the model converges. Despite ensuring that data stays within each owner’s domain, FedAvg has shown vulnerabilities on the security front, as it can be exposed to various types of attacks such as reconstruction attacks, poisoning attacks, and Sybil attacks. Additionally, FedAvg is sensitive to non-iid heterogeneous data from clients, which can lead to notable performance drops.

To address these challenges, Privacy-Preserving Federated Learning (PPFL) \cite{Yin2021ACS} was developed with the aim of integrating extra privacy and encryption techniques to enhance the intrinsic security of traditional federated learning. The most widely adopted privacy-preserving method is differential privacy (DP) \cite{ji2014differentialprivacymachinelearning} \cite{9714350}, which works by adding noise (typically Gaussian) to the data to mitigate potential damage from data leaks. Other strategies involve combining FedAvg with methods like SMPC \cite{hosseini2022clusterbasedsecuremultiparty} and homomorphic encryption (HE) \cite{jin2024fedmlheefficienthomomorphicencryptionbasedprivacypreserving} \cite{viand2023verifiablefullyhomomorphicencryption}.
Another significant issue highlighted by FedAvg concerns the management of non-identically distributed data among clients: in the presence of heterogeneity in data distribution among participants, the performance of the global model declines exponentially. Several recent studies aim to address the problem of non-IID data in a federated context: FedProx \cite{li2020federatedoptimizationheterogeneousnetworks} addresses the issue of non-IID data by adding a proximity term to the local loss function, which limits how much client models can deviate from the global model, thereby stabilizing training. SCAFFOLD \cite{karimireddy2021scaffoldstochasticcontrolledaveraging} introduces a correction for client drift through the use of variance reduction in gradients, ensuring more consistent updates among clients and accelerating convergence. MOON \cite{li2021modelcontrastivefederatedlearning}, on the other hand, leverages contrastive learning to align local models with the global model, reducing differences between local data distributions and improving generalization. FedDyn \cite{acar2021federatedlearningbaseddynamic} employs dynamic regularization that balances updates between clients and the central model, compensating for the bias introduced by the heterogeneous data distribution.

%\subsection{Our Contribute}

To mitigate the issue of heterogeneous data distribution among participants in a federated context, while ensuring the system remains scalable and transparent, we propose FedBGS: a blockchain-based self-segmented gossip learning framework designed to be scalable, model-agnostic, and resilient to non-IID data distributions. FedBGS is compatible with any Ethereum-based blockchain and, by leveraging IPFS, enables optimized data storage that would otherwise be unfeasible on public blockchains.

The main contributions of this work are as follows:

\begin{itemize}
    \item Federated K-Means++ for Automatic Segmentation: This work introduces a fully decentralized clustering mechanism using federated analytics, enabling automatic segmentation of clients based on data distributions and mitigating the impact of non-IID data.
    
    \item Blockchain-based Segmented Gossip Learning Approach: A blockchain-based segmented gossip learning approach is proposed, where clients exchange model updates within dynamically formed segments, ensuring scalability, robustness, and decentralization.
    
    \item Hybrid Blockchain and IPFS for Efficient Storage: The system integrates on-chain aggregation with off-chain storage via IPFS, overcoming blockchain scalability issues while maintaining transparency and security in decentralized learning.
    
    \item Empirical Validation on Non-IID Data: Extensive experiments on \textit{cifar-10} \cite{lecun1998gradient} and \textit{mnist} \cite{xiao2017fashion} \cite{krizhevsky2009learning} \cite{clanuwat2018kmnist} \cite{cohen2017emnist} datasets demonstrate improved security and accuracy compared to traditional federated learning and gossip learning methods in heterogeneous environments and resource-constrained blockchains. Furthermore, theoretical evaluations (Table \ref{tab:comparison}) will also be performed on FedBGS complete decentralization and scalability, which represent its true strong point compared to state-of-the-art methods.
\end{itemize}

\section{Related Works}

Despite the distributed nature of training, traditional federated architectures rely on centralized aggregation, making the server a critical and vulnerable point for the entire system. Several recent studies have focused on eliminating the server as a single point of failure (SPOF) in favor of peer-to-peer architectures, where aggregation occurs at the client level or via smart contracts leveraging blockchain technology. Frameworks such as FLchain \cite{FLchain}, BFLC \cite{BFLC}, and BlockFL \cite{kim2019blockchainedondevicefederatedlearning} use blockchain to ensure full decentralization in federated learning, providing participants with complete transparency over aggregation operations and the overall system context. However, similar to traditional federated learning, these frameworks still suffer from performance degradation in the presence of non-IID data, as well as scalability issues when using public blockchains, which are unsuitable for storing large amounts of model parameters. To address the storage limitations inherent in blockchain systems,\cite{8609675} proposed a hybrid architecture that combines on-chain and off-chain storage solutions. In this approach, the InterPlanetary File System (IPFS)\cite{benet2014ipfscontentaddressed} \cite{9182705} \cite{zhang2024decentralized} is utilized to store intermediate model parameters, thus mitigating the storage constraints imposed by blockchain bottlenecks.

Gossip learning \cite{Hegeds2019GossipLA} is a decentralized federated approach in which participants perform on-edge aggregation through a peer-to-peer mechanism, directly exchanging their model weights with other nodes in the network. This methodology overcomes the limitations of a central server, making the system more fault-tolerant and potentially more scalable. Empirical studies have demonstrated that gossip learning can achieve comparable, and in some cases superior, performance to federated learning, particularly when training data is evenly distributed across nodes.
Notably \cite{HEGEDUS2021109} provide an in-depth empirical comparison between gossip learning and federated learning, highlighting scenarios where the decentralized approach performs pairly or even exceeds centralized methods.

Split learning \cite{thapa2022splitfedfederatedlearningmeets} \cite{vepakomma2018splitlearninghealthdistributed} is a technique that reduces communication overhead by partitioning the neural network between the client and a central server. In this approach, clients compute the early layers locally and transmit only intermediate activations to the server, which then completes the forward and backward propagation. This strategy minimizes data exchange and enhances privacy by keeping raw data on the client; however, its reliance on a central server can reintroduce centralization issues and potentially limit scalability and security.

A significant extension of gossip learning is segmented gossip learning, which addresses the challenges posed by non-IID data distribution and communication overhead among nodes. Instead of sharing and updating the entire model at each step, only specific segments of the model parameters are exchanged among nodes. \cite{hu2019decentralizedfederatedlearningsegmented} discusses how to segment the model, typically the final layer, across clusters of nodes in a decentralized gossip learning framework. This approach reduces communication overhead and mitigates gradient conflicts in heterogeneous data scenarios, leading to faster convergence and improved model accuracy while maintaining a decentralized learning process.
In GossipFL \cite{9996127} , there is not an explicit segmentation but overhead is reduced by transmitting only the most significant model updates through sparsification, which minimizes the data exchanged among nodes. Additionally, adaptive communication dynamically adjusts the frequency and volume of updates based on network conditions and convergence flow, granting efficient resource utilization and faster training.

\begin{table*}[ht!]
\centering
\renewcommand{\arraystretch}{1.1}
\setlength{\tabcolsep}{6pt}
\caption{Comparison of FedBGS vs. Federated Learning (FL) and Blockchain-FL Approaches}
\label{tab:comparison}
\resizebox{\textwidth}{!}{%
\begin{tabular}{p{3cm} p{3.5cm} p{3.2cm} p{3.5cm} p{2.5cm} p{2.8cm}}
\toprule
\textbf{Approach} & \textbf{Non-IID Data Handling} & \textbf{Hybrid Storage (Blockchain + IPFS)} & \textbf{Privacy-Preserving Techniques} & \textbf{Scalability} & \textbf{Full Decentralization} \\
\midrule
FedAvg \cite{mcmahan2023communicationefficientlearningdeepnetworks} & Limited, requires enhancements & \texttimes & Basic privacy, vulnerable to inference attacks & High, but centralized & \texttimes \\
FLchain \cite{FLchain} & Moderate, inherits FedAvg limitations & Partial (On-chain only) & Transparency-focused, no additional privacy (DP, HE) & Moderate, consensus bottleneck & \checkmark \\
BFLC \cite{BFLC} & Limited, no special handling & Limited, On-chain with committee consensus & Robustness via committee validation & Improved but limited at scale & \checkmark \\
BlockFL \cite{kim2019blockchainedondevicefederatedlearning} & No special handling, standard FedAvg & On-chain only, high storage overhead & Transparent validation, no DP or encryption & Limited, consensus overhead & \checkmark \\
BGFL \cite{10127450} & Limited explicit non-IID handling, focuses on gossiping & On-chain only, no IPFS & Decentralized, no explicit DP or encryption & High via gossip communication & \checkmark \\
\textbf{FedBGS (Proposed)} & \textbf{Robust (One-Shot K-Means++ clustering)} & \textbf{Hybrid (Blockchain + IPFS)} & \textbf{HE, DP and Trimmed-Mean Aggregation} & \textbf{High, decentralized segmented training} & \textbf{\checkmark} \\
\bottomrule
\end{tabular}}

\vspace{0.1cm}
\noindent
\textbf{Note:} The bold entries indicate the notable strengths of FedBGS compared to other existing approaches.

\vspace{-0.3cm}
\end{table*}

\section{The Proposed Framework}
\begin{comment}
In a highly decentralized context, different clients can engage in shared training only through the use of federated learning, which addresses all the legal and economic issues that might arise from data sharing among parties. Gossip learning serves as an alternative to classical federated learning by removing the central server in favor of a peer-to-peer architecture capable of fully decentralized training, with parameter aggregation occurring directly on the client side.
\end{comment} 
\textbf{FedBGS} (Federated Blockchain Gossip Segmented) is a gossip learning based framework in which peer models are automatically segmented via Blockchain-Scheduling through One-Shot Federated K-means. This segmentation helps protect the system when client data distributions are heterogeneous and to avoid computational overhead. This approach allows handling non-iid data distributions while still providing each component with an overall view of the system. To ensure total transparency and trust among the parties involved (particularly relevant in cross-silo scenarios), FedBGS integrates a blockchain system based on Ethereum smart contracts. This makes it possible to validate participants, preventing malicious clients from infiltrating the process, and makes the system resistant to so-called “poisoning attacks” through dual validation at both the peer and blockchain levels. During the clustering phase, the blockchain also acts as a central aggregator, enabling the system to be fully decentralized and avoiding dependency on a single server that could become a SPOF. The limitations inherent in using Ethereum - a standard blockchain that is highly secure but offers limited scalability, is inefficient for on-chain storage of model weights, and has high gas fees - are addressed by a hybrid approach in which aggregation and storage are handled off-chain via IPFS, while a reference to the completed transaction is saved on-chain, along with the entire history of the federated process.

\subsection{\textit{Phase 1:} Auto-Clustering through Federated Analytics \eqref{alg:FedBGS_Phase1}} 

With the emergence of federated learning, another decentralized and privacy-oriented approach named Federated Analytics (FA) \cite{elkordy2023federatedanalyticssurvey}, began to take shape, of which federated learning is a subcategory. FA differs from classical federated learning because it is not aimed at training a global AI model, but rather at performing simple statistical calculations in a federated manner. Consequently, it has a much simpler structure and does not rely on multiple rounds. One-Shot Federated K-Means is a specific case of Federated Analytics in which the centroids are derived from the peers in a single round of communication. This makes it possible to group peers into clusters based on their similarity, without revealing each participant’s internal data. As in traditional federated structures, federated analytics also depends on the presence of a central aggregator. The advantage of this approach, however, lies in the fact that it does not require the sharing of full deep learning models. Simply sending statistical parameters, such as centroids, which require vastly less storage than weights, enables the use of smart contracts for on-chain storage and aggregation without incurring excessive fees.
It is worth noting that the clustering step was executed using a one-shot method typical of federated analytics, chosen specifically to avoid overloading the blockchain. On the flip side, one-shot methods often compromise clustering accuracy, typically settling for suboptimal solutions. However, our experiments revealed that this slight performance drop is generally negligible, even when employing a suboptimal approach. That said, few-shot iterative methods, which fall under the federated learning umbrella, can still be considered a viable option if their blockchain storage trade-offs are carefully assessed. To boost the overall precision of the clustering process, we employed the K-Means++ algorithm \cite{arthur2007k}, which selects the first centroid randomly and then chooses subsequent centroids based on a probability distribution proportional to the squared distance from the nearest existing centroid. This approach reduces randomness and improves clustering efficiency. FedBGS will use the clusters generated in the initial phase to create segments of the global model; each cluster corresponds to a segment derived from the model, meaning each cluster trains a different part of the model. The segmentation is done by splitting the model’s last layer into equal parts, ensuring that each cluster trains a specific portion of the model. This approach reduces gradient conflicts as well as computational and communication overhead.

The retrieval process works by accessing an on-chain mapping that associates each cluster with a specific range of neurons in the last fully connected layer. When a peer requests its segment boundaries, the smart contract fetches the assigned start and end indices from storage and returns them. This ensures that each cluster updates only the designated part of the model, preventing unauthorized modifications and enforcing decentralized segmentation.

Let \( C_k \) be the \( k \)-th cluster and \( \mathcal{S}_k \) denote its assigned segment of the model, defined by the neuron index range \([s_k, e_k]\). For a model parameter \( w_i \), the update rule enforced by the smart contract is:

\begin{align}
\label{eq:segment_restriction}
\mathcal{S}_k &= \{ w_i \mid s_k \leq i \leq e_k \} \\
\forall w_i \in \mathcal{W}, & \quad w_i \text{ is updated only if } w_i \in \mathcal{S}_k
\end{align}

\textbf{Legend:}\\
\( C_k \): The \( k \)-th cluster.\\
\( \mathcal{S}_k \): The segment assigned to cluster \( C_k \).\\
\( s_k, e_k \): Start and end indices defining the segment for \( C_k \).\\
\( w_i \): A model parameter in the last fully connected layer.\\
\( \mathcal{W} \): The set of all model parameters in the last layer.\\

\subsection{Blockchain Integration}
\begin{figure}[htbp]
    \centering
    \includegraphics{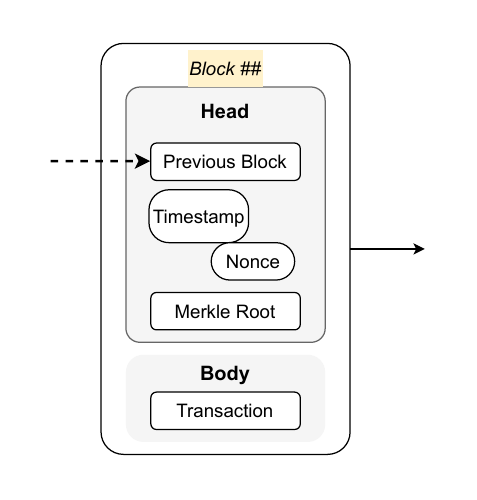}
    \vspace{-3pt} % Adjust the value as needed
    \caption{Example of an Ethereum Block: In Ethereum blocks, the Merkle root \cite{wikipediaMerkleTree} is a single hash that summarizes and verifies all transactions in the block, ensuring data integrity. The nonce is a number used in the consensus process to achieve a valid block hash. The previous block hash links the current block to the one before it, creating an immutable chain.}
    \label{fig:ethblock}
\end{figure}

Ethereum is a popular blockchain platform that lets people build and run smart contracts and decentralized applications without a central authority. The blockchain is made up of blocks, where each block (single block structure is shown in Fig. \ref{fig:ethblock}) is a list of transactions and a reference to the previous block. This linking of blocks creates a secure chain that makes it hard to change any past data.

In Ethereum, transactions are verified and new blocks are created by network participants. In the original system, these participants were called miners, and they used a method called Proof-of-Work (PoW) to solve complex puzzles before adding a block. However, Ethereum has now moved to a system called Proof-of-Stake (PoS). In PoS, instead of miners, there are validators who are chosen to create blocks based on the amount of Ether they lock up (or “stake"). This method uses much less energy and is more efficient.

Modern validation techniques in Ethereum focus on making sure that validators act honestly. For example, if a validator behaves badly, the system can “slash" (or penalize) their stake. This helps keep the network secure and reliable.

We chose to use Ethereum for our project on federated gossip learning and as a server for federated one-shot K-means because it offers a decentralized and transparent way to verify updates and client registrations. By using Ethereum, we can ensure that every participant's contribution is recorded in a secure, tamper-proof ledger.

In Decentralized learning, blockchain is typically used for the following purposes:
\begin{itemize}
    \item \textbf{Decentralization:} Decentralizing the system by performing on-chain aggregation via smart contracts, thus eliminating the server as a single point of failure.
    \item \textbf{Immutable Client Registry:} Maintaining an immutable ledger of clients authorized to participate, which helps prevent Sybil attacks.
    \item \textbf{Secure Transaction Recording:} Ensuring that transactions (i.e., updates sent by nodes) are recorded in a distributed and unchangeable manner, thereby reducing the risk of tampering (poisoning attacks).
    \item \textbf{Double Spending Prevention:} Preventing double spending attacks, where a node might try to reuse old updates or replicate them as new to manipulate the federated model, since every transaction is uniquely tracked and cannot be “spent" again.
    \item \textbf{Incentivization:} Implementing incentive systems through token-based smart contracts.
\end{itemize}

\subsection{IPFS and Blockchain}
\begin{figure}[htbp]
    \centering
    \includegraphics[width=0.5\textwidth]{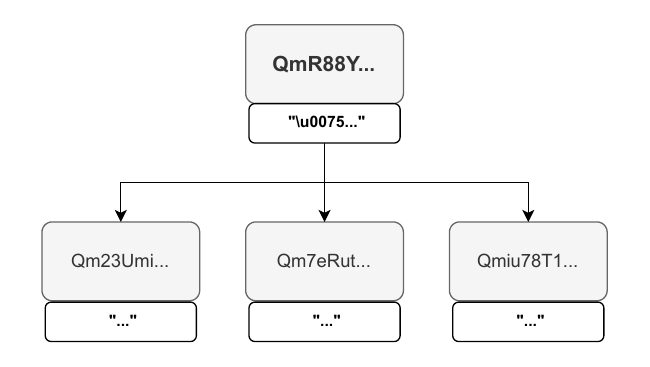}
    \vspace{-5pt} % Adjust vertical spacing as needed
    \caption{The block size in IPFS can vary depending on the stored data. Thanks to a hierarchical structure based on hash links, each block points to other linked blocks, promoting redundancy and sharing across nodes. This scalable approach ensures fast and reliable retrieval of content from various locations.}
    \label{fig:ipfs}
\end{figure}

IPFS (InterPlanetary File System) is a system for storing and sharing files in a distributed manner, without relying on centralized servers. The basic principles of IPFS are as follows:

\begin{itemize}
    \item \textbf{Content Addressing:}\\
    In IPFS, files are not identified by their location (such as a URL), but by their content as show in Fig. \ref{fig:ipfs}. When you add a file, it is transformed into a unique cryptographic hash. This hash becomes the \emph{address} of the file, ensuring that each file is securely and immutably identified.
    
    \item \textbf{Blocks and the Merkle DAG:}\\
    A file in IPFS is divided into small pieces called \emph{blocks}. Each block contains:
    \begin{itemize}
        \item \textbf{Data:} A portion of the file.
        \item \textbf{Links:} References to other blocks (if the file is split into multiple parts).
    \end{itemize}
    These links form a structure called a \emph{Merkle DAG} (Directed Acyclic Graph), which allows verification that each block has not been altered. Even if a single block changes, the overall hash will change, indicating a modification.
    
    \item \textbf{Distribution:}\\
    When you request a file by its hash, IPFS searches the entire network for the block (or blocks) corresponding to that hash and reassembles them to reconstruct the file.
\end{itemize}

\begin{comment}
\begin{figure}[htbp]
    \centering
    \begin{Verbatim}[frame=single, fontsize=\small]
fabio@fabio-VBox:~/tmp$ ipfs add test_dir/...
added QmR88YmkVV...wuG9w test_dir/bigfile.js
fabio@fabio-VBox:~/tmp$ ipfs object get QmR45
{
  "Links": [
    {
      "Name": "",
      "Hash": "Qm23UmiyM3...Ydk8DQ6KKC",
      "Size": 262158
    },
    {
      "Name": "",
      "Hash": "Qm7eRutjFm...b4EQdfWz54",
      "Size": 262158
    },
    {
      "Name": "",
      "Hash": "Qmiu78T1hj...U5mfUb3sVH",
      "Size": 178947
    }
  ],
  "Data": "\u0075\u3333\u00...u0345\n"
}
    \end{Verbatim}
    \caption{In this example, the blocks do not have a name because they are uniquely identified by their hash, ensuring immutability and traceability of the data. The interaction between IPFS and Ethereum is advantageous, as it enables decentralized off-chain storage, reducing costs and enhancing system resilience.}
    \label{fig:ipfs_code}
\end{figure}
\end{comment}

Both IPFS and Ethereum utilize a Merkle graph structure to efficiently ensure data integrity and verify content with minimal cryptographic proofs. In Ethereum, this design (Patricia-Tree) enables fast, decentralized verification of transactions and state changes, reducing gas costs and enhancing scalability. Similarly, IPFS’s use of a Merkle graph allows for immutable, content-addressed storage, creating a natural synergy that makes integration between the two systems both seamless and robust.

With FedBGS, the use of the blockchain will be divided into two phases:

\textbf{Phase 1:}
\begin{itemize}
    \item \textbf{Task 1:} The blockchain will have the role of managing the registration of each participant through unique credentials; an unregistered client will not have the right to participate in the shared training.
    \item \textbf{Task 2:} We will perform the aggregation of the centroids for the federated k-means via smart contracts and store the results directly on-chain.
    \item \textbf{Task 3:} The smart contract, using an internal policy algorithm, determines which part of the model will be assigned to each cluster. At the start of the gossip cycle, the client queries the smart contract to learn its designated cluster and which “segment” of the model it should train.
\end{itemize}

\textbf{Phase 2:}
\begin{itemize}
    \item \textbf{Task 1:} At regular intervals, the smart contract will randomly select a leader to handle global inter-segment aggregation by regenerating the original, “non-segmented” model, which is then stored on IPFS.
    \item \textbf{Task 2:} When each peer retrieves the global model, a validation process is carried out to verify the model’s integrity (the hash of the model extracted by the peer must match the immutable one stored on the blockchain).
    \item \textbf{Task 3:} A token-based incentive system rewards clients who provide a better model to the subsequent client, penalizing those who deviate too much from the current loss.
\end{itemize}

\subsection{Privacy-Preserving Decentralized Learning}

As mentioned in the introduction, the intrinsic security level provided by gossip learning is not sufficient to ensure an adequate level of privacy, especially in cross-silo contexts. The introduction of blockchain with IPFS into the system aims to prevent some possible attacks such as Sybil or poisoning attacks; data leakage or the potential espionage by “curious" participants remains an unresolved issue with the previously described technologies, particularly during the clustering phase, where the transmission of statistical data such as centroids rather than parameters of a deep model makes the system especially vulnerable to several types of attacks, including:

\begin{itemize}
    \item \textbf{Membership Inference Attacks:}  
    An adversary attempts to determine whether a particular sample (e.g., a user's data) was present in the training set.
    
    \item \textbf{Gradient Inversion (or Reconstruction) Attacks:}  
    The attacker tries to reconstruct sensitive original data (images, text, etc.) from the gradients shared by clients during training.
    
    \item \textbf{Model Inversion Attacks:}  
    The goal is to infer sensitive attributes of the training examples by leveraging information returned by the model.
    
    \item \textbf{Eavesdropping / Man-in-the-Middle Attacks:}  
    An attacker intercepts or modifies parameters (or gradients) while they are transmitted between participants and the central server (or among peers).
    
    \item \textbf{Model Extraction Attacks:}  
    The adversary attempts to extract or replicate the model's architecture and parameters by making queries or analyzing the system's responses.
\end{itemize}

In FedBGS, we integrate several advanced techniques to enhance the security and robustness of our decentralized gossip learning system, which leverages blockchain and IPFS. These techniques, including Differential Privacy, Homomorphic Encryption, and robust aggregation via the trimmed mean, work together to protect sensitive information, ensure secure aggregation, and defend against adversarial attacks.

\textbf{Differential Privacy (DP):}  
DP ensures that the contribution of each client remains private, even when model updates are shared. In federated settings, \emph{local DP} involves adding noise at the client level before sending updates, whereas \emph{global DP} applies noise during the aggregation process. Within our gossip learning framework, DP is used during the training phase to clip and perturb gradients, thereby safeguarding individual data contributions while still enabling effective collaborative learning.

During the training phase in segmented gossip learning, each client's gradient \( g \) is made differentially private by clipping and adding Gaussian noise. This is represented by:
\begin{align}
\label{eq:dp}
\tilde{g} = g \cdot \min\!\left(1, \frac{C}{\|g\|_2}\right) + \mathcal{N}\!\left(0, \sigma^2 C^2\right)
\end{align}
\textbf{Legend:}\\
\( g \): Original gradient computed by the client.\\
\( \tilde{g} \): Differentially private gradient.\\
\( C \): Clipping threshold.\\
\( \|g\|_2 \): \( L_2 \)-norm of the gradient \( g \).\\
\( \mathcal{N}(0, \sigma^2 C^2) \): Gaussian noise with mean 0 and variance \( \sigma^2 C^2 \).\\
\( \sigma \): Noise multiplier controlling the amount of noise added.

\textbf{FedBGS} implements an internal scheduler that allows DP to be changed dynamically. Our scheduler applies a linear decrease to the DP over time, reaching the minimum level in the final iteration. The early rounds are particularly delicate because the model is directly influenced by each node's initial data. Since the model has not yet been “mixed" with information from many other participants, the updates sent may contain more identifiable traces of the original data. In other words, privacy could be at greater risk because the “noise" (i.e. the variability introduced by other nodes and any artificial DP noise) has not yet accumulated enough to mask the individual characteristics of the local datasets.

Furthermore, in the initial rounds, if the updates are more significant, it is necessary to introduce more noise to keep them protected, which can quickly consume the privacy budget. If this budget is exhausted or significantly reduced in the very early rounds, there is a risk of not being able to guarantee the same level of protection in subsequent iterations.

\textbf{Homomorphic Encryption (HE):}  
HE allows computations to be carried out on encrypted data without revealing the underlying information. Although \emph{fully homomorphic encryption} supports arbitrary computations on ciphertexts, its high computational cost makes it less practical. Instead, we employ \emph{partial homomorphic encryption}, specifically, the additively homomorphic Paillier scheme to securely aggregate label distributions and model updates. This enables us to combine encrypted client updates without exposing their raw values, maintaining both privacy and integrity during the aggregation process.

To protect the centroids during the clustering phase, we aggregate the encrypted label distributions using the Paillier scheme \cite{cao2015paillierscryptosystemvariantsrevisited}. Let \( x_{i,j} \) be the \( j \)-th component of the label distribution (centroid) from client \( i \). With \( E(\cdot) \) and \( D(\cdot) \) denoting the encryption and decryption functions, respectively, and using the homomorphic addition operator \( \bigoplus \), the aggregated (and decrypted) centroid component is computed as:
\begin{align}
\label{eq:HE}
\hat{c}_j = \frac{D\!\left( \bigoplus_{i=1}^{n} E\left(x_{i,j}\right) \right)}{n}
\end{align}
\textbf{Legend:}\\
\( E(x) \): Encryption of \( x \) using the Paillier scheme.\\
\( D(y) \): Decryption of ciphertext \( y \).\\
\( \bigoplus \): Homomorphic addition operator.\\
\( n \): Total number of clients (or label distributions).\\
\( x_{i,j} \): \( j \)-th component of client \( i \)'s label distribution.

In \textbf{FedBGS}, partial homomorphic encryption is used only during the clustering phase, which we identify as the most vulnerable from a security standpoint. It is not applied throughout the entire process to avoid the computational overhead caused by gossiping (more data to encrypt and more iterations).

\textbf{Trimmed Mean Aggregation:}  
To counteract Byzantine attacks \cite{cajaravilleaboy2024byzantinerobustaggregationsecuringdecentralized}, where some nodes might submit malicious updates, aggregation is computed utilizing the trimmed mean. This robust statistical method discards a fixed proportion of the most extreme values before averaging, thereby reducing the impact of outliers. In the context of our gossip learning system, the trimmed mean helps maintain the reliability of the global model by mitigating the effects of adversarial contributions.

Together, these techniques enable our system to perform secure, privacy-preserving, and robust model aggregation in a federated environment. By combining DP for privacy, partial HE for secure computations, and trimmed mean for robust aggregation, our approach is well-suited to address the unique challenges posed by decentralized and potentially adversarial settings.

Let \( \{\theta_i\}_{i=1}^{n} \) be the set of model updates and \( \theta_{(i)} \) denote the \( i \)-th sorted update. With a trim ratio \( r \), the robust aggregated update \( \theta^* \) is computed as:
\begin{align}
\label{eq:trim_mean}
\theta^* = \frac{1}{n(1-2r)} \sum_{i=\lceil rn \rceil}^{\lfloor (1-r)n \rfloor} \theta_{(i)}
\end{align}
\textbf{Legend:}\\
\( \theta_i \): The \( i \)-th individual model update.\\
\( \theta_{(i)} \): The \( i \)-th order statistic (sorted update).\\
\( n \): Total number of updates.\\
\( r \): Trim ratio (fraction of values trimmed from each end).\\
\( \theta^* \): The robust aggregated model update.

\begin{algorithm}[!t]
\caption{FedBGS Phase 1: On-chain Clustering \& Client Registration}
\label{alg:FedBGS_Phase1}
\begin{algorithmic}[1]
\Require Peer set \(P\) with local data \(D_p\)
\Ensure Each peer \(p \in P\) is registered on-chain and assigned to a segment \(S_i\)
\Statex
\For{each peer \(p \in P\)}
    \State Register \(p\) on the blockchain with unique credentials
    \State Compute local label distribution \(x_p\) from \(D_p\)
\EndFor
\State Form segments \(S_1, S_2, \dots, S_S\) via federated k-means++ on \(\{x_p\}_{p \in P}\)
\For{each segment \(S_i\)}
    \State \textbf{Secure Aggregation:} Aggregate label distributions using Partial HE
    \Comment{see HE formulation in Eq.~\eqref{eq:HE}}
    \State Compute secure centroid \(\hat{c}_i\) for segment \(S_i\)
    \State Update \(\hat{c}_i\) on-chain (acting as the aggregation server)
\EndFor
\State \textbf{Cluster Discovery:} Each peer \(p \in P\) retrieves its assigned segment
\Statex \quad \(S_i\) via smart-contract query subject to Eq.~\eqref{eq:segment_restriction}
\end{algorithmic}
\end{algorithm}

\begin{algorithm}[!t]
\caption{FedBGS Phase 2: Segmented Gossip Learning with Hybrid Storage \& Global Aggregation}
\label{alg:FedBGS_Phase2}
\begin{algorithmic}[1]
\Require Peer set \(P\) with assigned segments \(S_i\), DP parameters \(C,\,\sigma\), trim ratio \(r\)
\Ensure Global aggregated model parameters \(\theta^*\)
\Statex
\For{each peer \(p \in P\) within its assigned segment \(S_i\)}
    \State Train local model and compute gradient \(g_p\)
    \State \textbf{Apply Differential Privacy:} Clip and perturb \(g_p\) to obtain \(\tilde{g}_p\)
    \Comment{see DP formulation in Eq.~\eqref{eq:dp}}
    \State Share \(\tilde{g}_p\) via hybrid storage
    \Statex \quad -- Store model update on IPFS
    \Statex \quad -- Record update hash on-chain for integrity verification
    \State Validate the update by matching the IPFS hash with the blockchain record
    \If{validation fails}
        \State Penalize \(p\) via smart-contract enforced token deduction
    \EndIf
\EndFor
\State \textbf{Global Aggregation:}
\State Elect a leader \(p^*\) via smart contract (e.g., using on-chain randomness)
\For{each segment \(S_i\)}
    \State Leader \(p^*\) collects updates \(\{\tilde{g}_p\}_{p \in S_i}\)
    \State Aggregate updates using trimmed mean (Eq.~\eqref{eq:trim_mean})
    \State Obtain segment update \(\Delta \theta_i\)
    \State Update global model segment \(i\) with \(\Delta \theta_i\)
\EndFor
\State Leader \(p^*\) reconstructs the complete global model \(\theta^*\)
\State Store \(\theta^*\) on IPFS and record its CID on-chain
\State Each peer \(p \in P\) retrieves the latest CID and updates its local model
\end{algorithmic}
\end{algorithm}

\subsection{\textit{Phase 2: }Segmented Gossip Learning \eqref{alg:FedBGS_Phase2}}

As mentioned earlier, gossip learning (GL) is a strategy that allows training a shared model in a completely decentralized manner, without participants explicitly sharing their data. Gossip learning differs from federated learning (which is semi-decentralized) in that it does not rely on a central server for aggregation; instead, aggregation is performed at the client level, making the system immune to malicious or faulty servers.

Gossip learning can operate under two different paradigms:

Synchronous Logic: This approach retains the concept of rounds as in federated learning, where all peers wait for every participant to complete the training phase before sending their model updates. Although this method is simpler and more orderly from an analytical standpoint, it creates significant bottlenecks when clients are heterogeneous, making the system less scalable in scenarios with many participants having varying computational capabilities and data.
Asynchronous Logic: In this approach there are no rounds; each client aggregates model updates at different times based on its internal policies. While this strategy is less orderly, it offers better performance compared to the synchronous method.

\begin{figure*}[htb]
  \centering
  % Prima immagine con didascalia dedicata
  \subfloat[\textbf{Phase 1}: Each client is registered on the blockchain and clustered based on data similarity. The blockchain serves as an aggregator for centroid updates, leveraging a federated k-means clustering approach. The FedClustering operates in a decentralized manner using a One-Shot approach, where the centroids are stored on-chain. Cluster assignments directly influence model segmentation, as each cluster is responsible for a subset of the model params.]{%
    \includegraphics[width=0.48\textwidth]{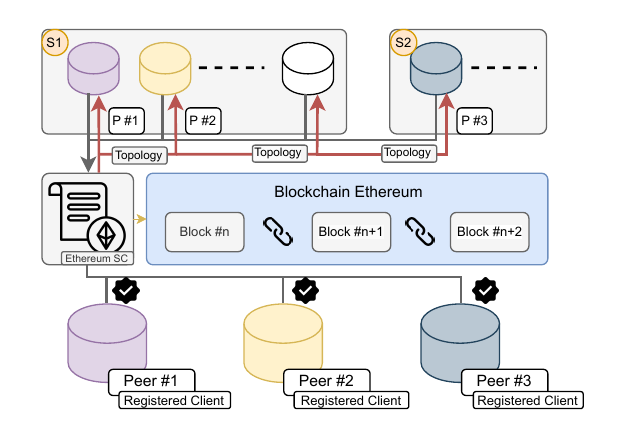}%
    \label{subfig:phase1}
  }
  \hfill
  % Seconda immagine con didascalia dedicata
  \subfloat[\textbf{Phase 2}: Clients perform differentially private cluster-based segmented gossip learning, where only specific segments of the last fully connected layer are updated by each cluster. Model updates are stored and retrieved from IPFS, ensuring efficient validation and learning transparency. Aggregation on the client side is performed using trimmed-mean aggregation, and the incentive system for peer training is managed through a dedicated smart contract.]{%
    \includegraphics[width=0.48\textwidth]{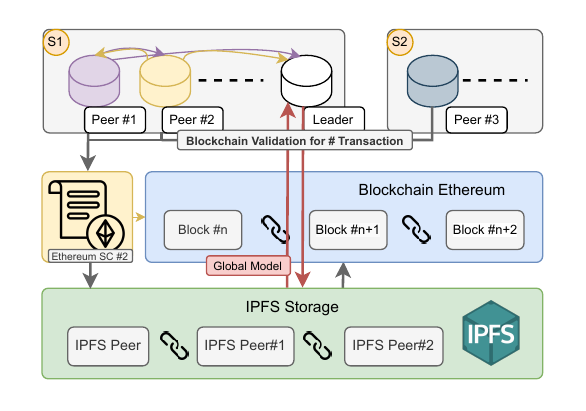}%
    \label{subfig:phase2}
  }

  % Didascalia generale
  \caption{FedBGS overall architecture with cluster-based segmented gossip learning. Thanks to smart contracts and Ethereum's execution logic, every operation delegated to the blockchain is executed across multiple blocks and validated. The segmentation strategy ensures efficient training on non-IID data by limiting updates to assigned neuron blocks in the final layer.}
  \label{fig:combined}
\end{figure*}

As outlined in the overall architecture shown in Fig. \ref{fig:combined}, FedBGL operates in two phases:

\textbf{Phase 1 (Fig. \ref{subfig:phase1}): On-Chain Clustering \& Client Registration}: 
\begin{itemize}
    \item \textbf{Client Registration:} Each peer registers on the blockchain using unique credentials.
    \item \textbf{Local Computation:} Every peer computes its own local label distribution from its local dataset.
    \item \textbf{Clustering:} A federated k-means++ algorithm is applied to these distributions to form homogeneous segments.
    \item \textbf{Secure Aggregation:} For each segment, the label distributions are aggregated securely using partial homomorphic encryption to compute a secure centroid.
    \item \textbf{On-Chain Update:} The secure centroids are updated directly on-chain, acting as the aggregation servers.
    \item \textbf{Segment Assignment:} Each peer retrieves its assigned segment \(S_i\) by querying the blockchain via the smart contract.
\end{itemize}

\textbf{Phase 2 (Fig. \ref{subfig:phase2}): Segmented Gossip Learning with Hybrid Storage \& Global Aggregation}: 
\begin{itemize}
    \item \textbf{Local Training:} Within each segment, peers train their local models and compute gradients.
    \item \textbf{Differential Privacy:} Each gradient is processed with differential privacy mechanisms to protect sensitive information.
    \item \textbf{Hybrid Storage:} The privacy-preserving updates are shared by:
        \begin{itemize}
            \item Storing the model update on IPFS.
            \item Recording the corresponding update hash (CID) on the blockchain for verification.
        \end{itemize}
    \item \textbf{Update Verification:} Each peer validates the update hash by comparing the IPFS hash with the blockchain record. In case of validation failure, the peer is penalized via blockchain token deduction.
    \item \textbf{Leader Election:} A leader peer is elected on-chain via a smart contract (e.g., using on-chain randomness).
    \item \textbf{Robust Aggregation:} The elected leader collects the privacy-preserving updates from peers in each segment and aggregates them using a trimmed mean approach to compute the segment update.
    \item \textbf{Global Model Reconstruction:} The leader reconstructs the complete global model from the aggregated segment updates.
    \item \textbf{Global Storage \& Dissemination:} 
        \begin{itemize}
            \item The leader stores the global model on IPFS and records the corresponding CID on-chain.
            \item Each peer retrieves the latest global model via the recorded CID and updates its local model segment accordingly.
        \end{itemize}
\end{itemize}

In FedBGS, in the second phase, we implemented an asynchronous gossip learning system where peers, operating independently (each in its own thread), perform local model training. Training, sharing, and testing operations occur asynchronously, with variable intervals determined randomly.

\section*{Global Model Derivation and Iteration of Gossip Learning}

In the proposed protocol, each peer begins by training its local model on its private dataset and computing a gradient update, which is then processed through Differential Privacy mechanisms to preserve data confidentiality. Within each homogeneous segment, obtained via a federated k-means++ clustering algorithm, peers exchange their privacy-preserving updates with their designated neighbors. The intra-segment aggregation, performed using a robust method such as the trimmed mean, refines the local model for that segment.

At regular intervals, the system elects a leader peer on-chain via a smart contract that employs a randomization mechanism. The elected leader collects the aggregated updates from all segments and fuses them into a unified global model. Subsequently, the leader stores the global model on IPFS and records the corresponding CID on the blockchain. Finally, each peer retrieves the updated global model using the recorded CID and updates its local model accordingly.

This iterative process of local training, intra-segment aggregation, leader-driven global aggregation, and model dissemination ensures continuous improvement and convergence of the global model.

\begin{comment}
\begin{figure*}[htb]
    \centering{
        \includegraphics[width=0.8\textwidth]{images/overall arch.pdf}
    }
    \caption{Overall FedBGS architecture illustrating a two-phase system for decentralized learning. In Phase 1, each client is registered on the blockchain and clustered by data similarity, with the blockchain acting as an aggregator for secure centroid updates. In Phase 2, clients perform segmented gossip learning, storing and retrieving model updates from IPFS for efficient client validation. Together, these phases ensure robust, privacy-preserving training in a fully decentralized environment.}
    \label{fig:fl_vs_all_vs_single}
\end{figure*}
\end{comment}

\section{Experimental Evaluation}

The objective of this experimental evaluation is to graphically demonstrate the convergence of FedBGS using various datasets and under different initial data distribution conditions. The contribution of individual participants to the training will be shown through line charts and bar plots, and a table summarizing the impact of FedBGS on an Ethereum blockchain in terms of costs will be included. Other evaluations related to the security and scalability of the system will be conducted at a theoretical level.
To validate the performance of FedBGS, we tested it on 6 different standard datasets.
This document provides an elegant overview of the standard datasets used to validate the performance of FedBGS. The datasets are grouped into two main categories: those based on grayscale images (28$\times$28) and the CIFAR-10 dataset, which uses color images (32$\times$32).
The following experiments were all executed locally on a NVIDIA RTX 4080 laptop GPU and are fully replicable by following the deployment instructions on GitHub\footnote{\url{https://github.com/FabioTur-dev/gossip_bc_full}}.

\subsection*{Grayscale Image Datasets (28$\times$28)}
These datasets are characterized by their uniform size and grayscale format. They are widely adopted as benchmarks due to their simplicity, standardized structure, and ease of comparison across models.

\begin{itemize}
    \item \textbf{MNIST:}\\
          Contains handwritten digits (0--9) represented as $28\times28$ grayscale images. It comprises 60,000 training images and 10,000 test images.
          
    \item \textbf{EMNIST:}\\
          An extension of MNIST that includes additional letter classes. The “digits" split mirrors MNIST, offering a larger dataset with approximately 280,000 images in total.
          
    \item \textbf{Fashion-MNIST (FMNIST):}\\
          Consists of $28\times28$ grayscale images of clothing items, with 60,000 training and 10,000 test images. It provides a more challenging alternative to digit recognition.
          
    \item \textbf{KMNIST:}\\
          Comprises $28\times28$ grayscale images of Japanese cursive (Kuzushiji) characters. It generally follows a similar split to MNIST (around 60,000 training and 10,000 test images), offering a unique visual challenge.
\end{itemize}

\subsection*{CIFAR-10}
\begin{itemize}
    \item \textbf{CIFAR-10:}\\
          Consists of $32\times32$ RGB color images across 10 classes (e.g., airplanes, cars, birds). It includes 50,000 training images and 10,000 test images, and its natural complexity and diversity make it a challenging benchmark for image classification tasks.
\end{itemize}

\subsection{Client Models}\label{AA}
For the experimental validation of the results, we implemented two different convolutional networks. The first (NetMNIST) is used generically for all grayscale datasets derived from MNIST; the second, on the other hand (NetCIFAR), is more complex and is used to achieve the desired performance with an RGB dataset such as CIFAR-10. Further experiments on larger datasets such as STL-10 and ImageNet were not conducted due to our hardware limitations.

\subsection*{NetMNIST}
The \textbf{NetMNIST} network is a CNN specifically designed for the MNIST dataset, which contains $28\times28$ grayscale images representing handwritten digits. Its architecture consists of:
\begin{itemize}
    \item \textbf{Convolutional Layers:} Two convolutional layers are used. The first layer employs 6 filters with a kernel size of 5, while the second layer uses 16 filters (also with a kernel size of 5). These layers, followed by ReLU activations and max pooling operations, extract the salient features from the images.
    \item \textbf{Flattening:} The output of the convolutional layers is flattened to prepare it for the subsequent layers.
    \item \textbf{Fully Connected Layers:} A sequence of three fully connected layers is used, with 120 and 84 neurons in the first two layers, respectively, before mapping the data into the 10 classes (from 0 to 9).
\end{itemize}
This simple and lightweight architecture is ideal for digit recognition due to the standardized and relatively simple nature of the MNIST dataset.

\subsection*{NetCIFAR (for CIFAR-10)}
The \textbf{NetCIFAR} network was developed for the CIFAR-10 dataset, which consists of $32\times32$ RGB color images distributed across 10 classes. Its architecture is notably more complex, featuring:
\begin{itemize}
    \item \textbf{Convolutional Structure:} A sequence of convolutional layers with $3\times3$ kernels is employed. The first block uses 32 filters, the second block uses 64 filters, and the third block utilizes 128 filters. Each convolutional layer is followed by batch normalization and ReLU activations to enhance training stability.
    \item \textbf{Pooling:} Two max pooling operations are applied, reducing the spatial dimensions of the feature maps to $8\times8$, which facilitates handling the increased image complexity.
    \item \textbf{Classifier:} After flattening the feature maps, the classifier includes a fully connected layer with 256 neurons, followed by dropout to mitigate overfitting, and finally a linear layer that maps the features into the 10 classes.
\end{itemize}
This higher number of filters and neurons highlights the increased complexity and diversity of the natural images in the CIFAR-10 dataset, ensuring a good balance between generalization capacity and training stability.

\subsection*{Dirichlet Partitioning}

In our setup, we partition the dataset among clients using a Dirichlet distribution \cite{osting2017consistencydirichletpartitions} \cite{li2021federatedlearningnoniiddata}. This approach simulates non-IID data by assigning different proportions of each class to different clients. For a dataset with $K$ classes and $N$ clients, we generate, for each class $k$, a probability vector 
\begin{align}
    \mathbf{p}_k \sim \mathrm{Dirichlet}(\beta, \beta, \ldots, \beta),
\end{align}
where $\beta$ is a concentration parameter that controls the degree of heterogeneity. A smaller $\beta$ produces a more imbalanced distribution, whereas a larger $\beta$ results in a more uniform, IID-like partitioning.

For each class $k$, given that there are $n_k$ samples, the number of samples allocated to client $i$ is calculated as:
\begin{align}
    n_{k,i} = \left\lfloor p_{k,i} \cdot n_k \right\rfloor,
\end{align}
where $p_{k,i}$ is the $i$-th component of $\mathbf{p}_k$. Any remaining samples are then randomly distributed among the clients.

This partitioning strategy is applied to simulate realistic data heterogeneity among clients, which is critical for evaluating the robustness and performance of our federated learning approach under non-IID conditions.

\begin{comment}
\begin{figure}[htbp]
\centerline{\includegraphics[width=0.55\textwidth]{images/communication.pdf}}
\caption{This communication graph shows the number of interactions between peers during a FedBGS training cycle. The two different colors on the nodes represent the peers belonging to their respective clusters (in this case, there are 2 clusters) following the segmentation phase.}
\label{fig::communication graph}
\end{figure}
\end{comment}

\begin{figure*}[htb]
  \centering
  % Prima riga: tre immagini (loss)
  \subfloat{\includegraphics[width=0.32\textwidth]{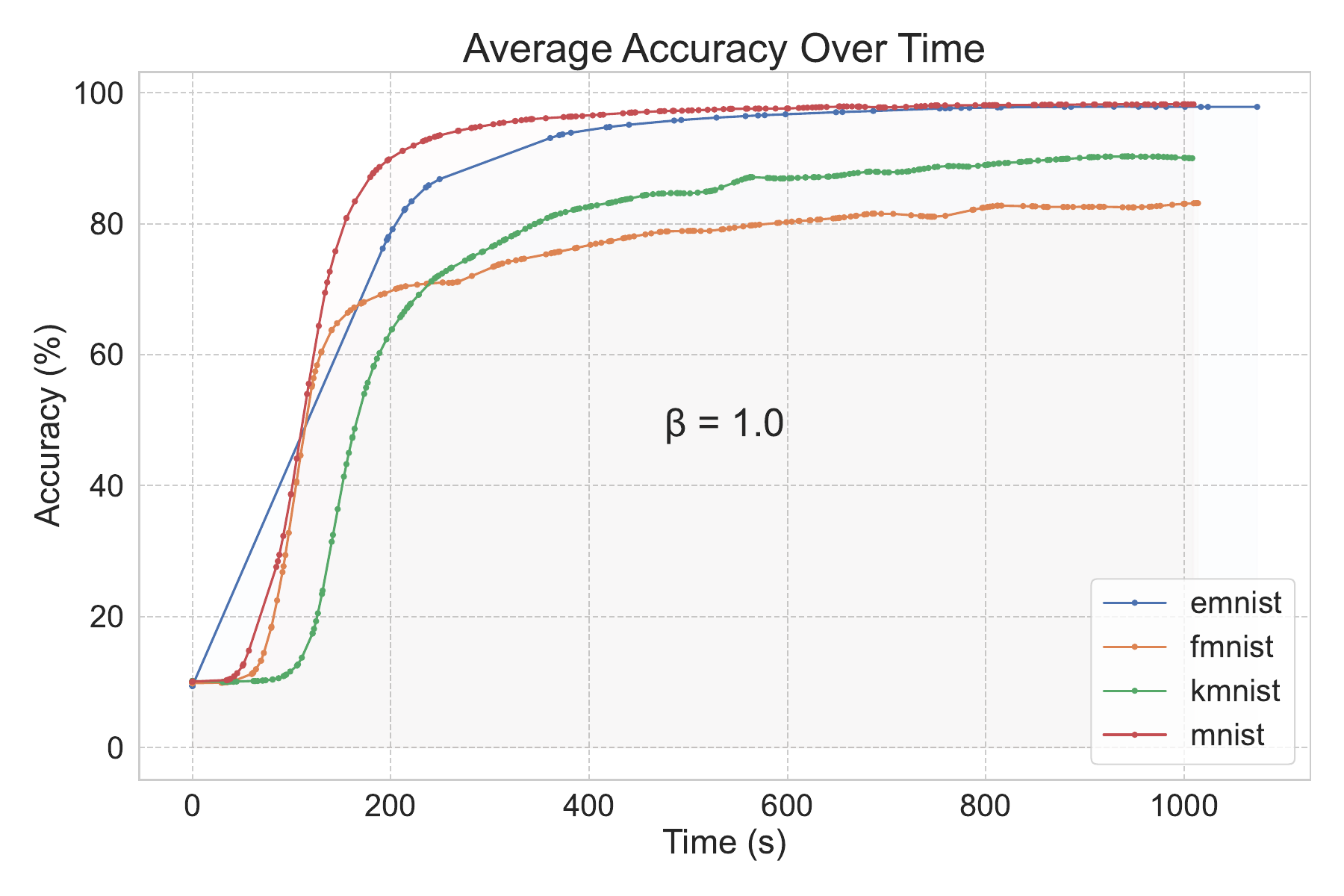}}\hfill
  \subfloat{\includegraphics[width=0.32\textwidth]{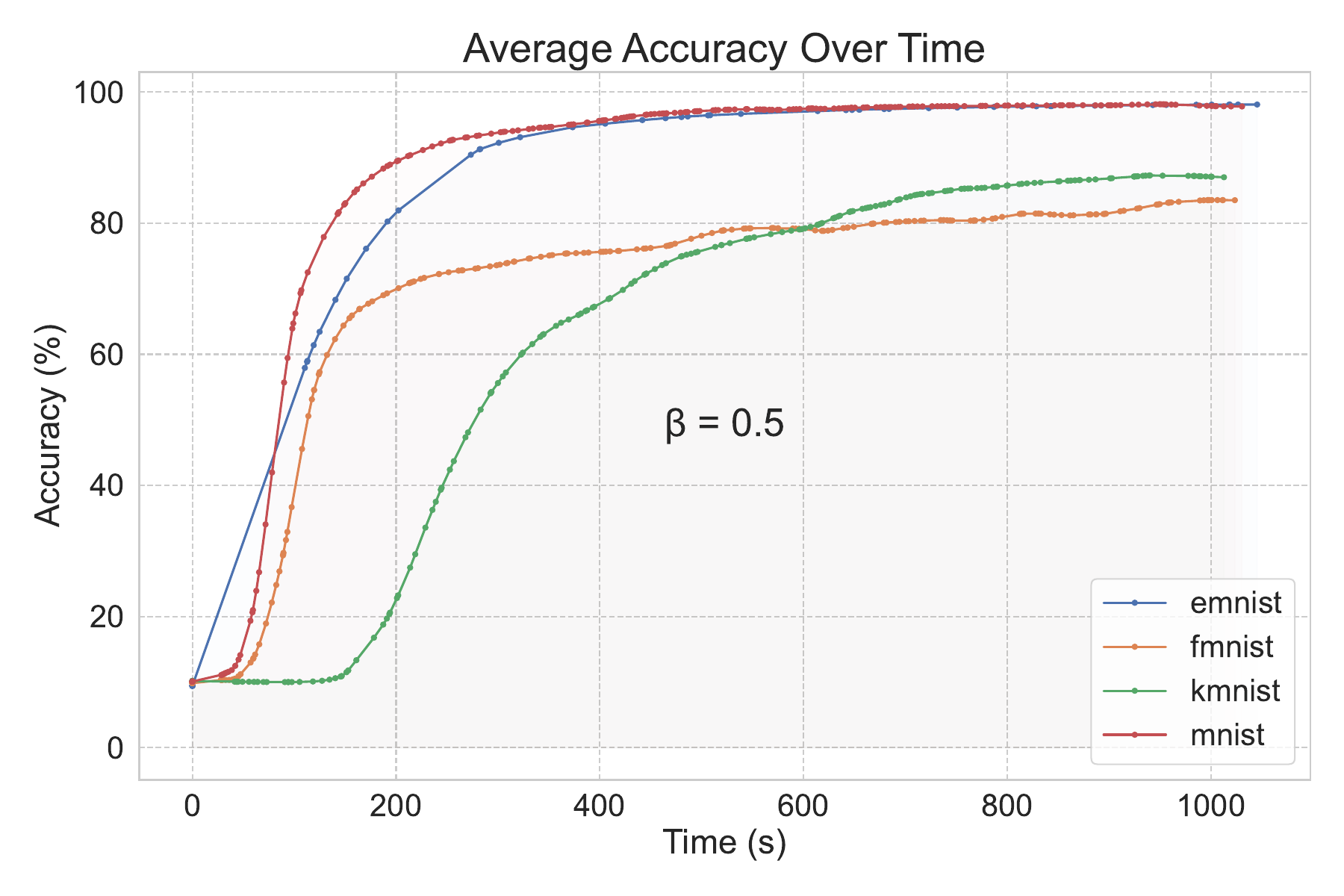}}\hfill
  \subfloat{\includegraphics[width=0.32\textwidth]{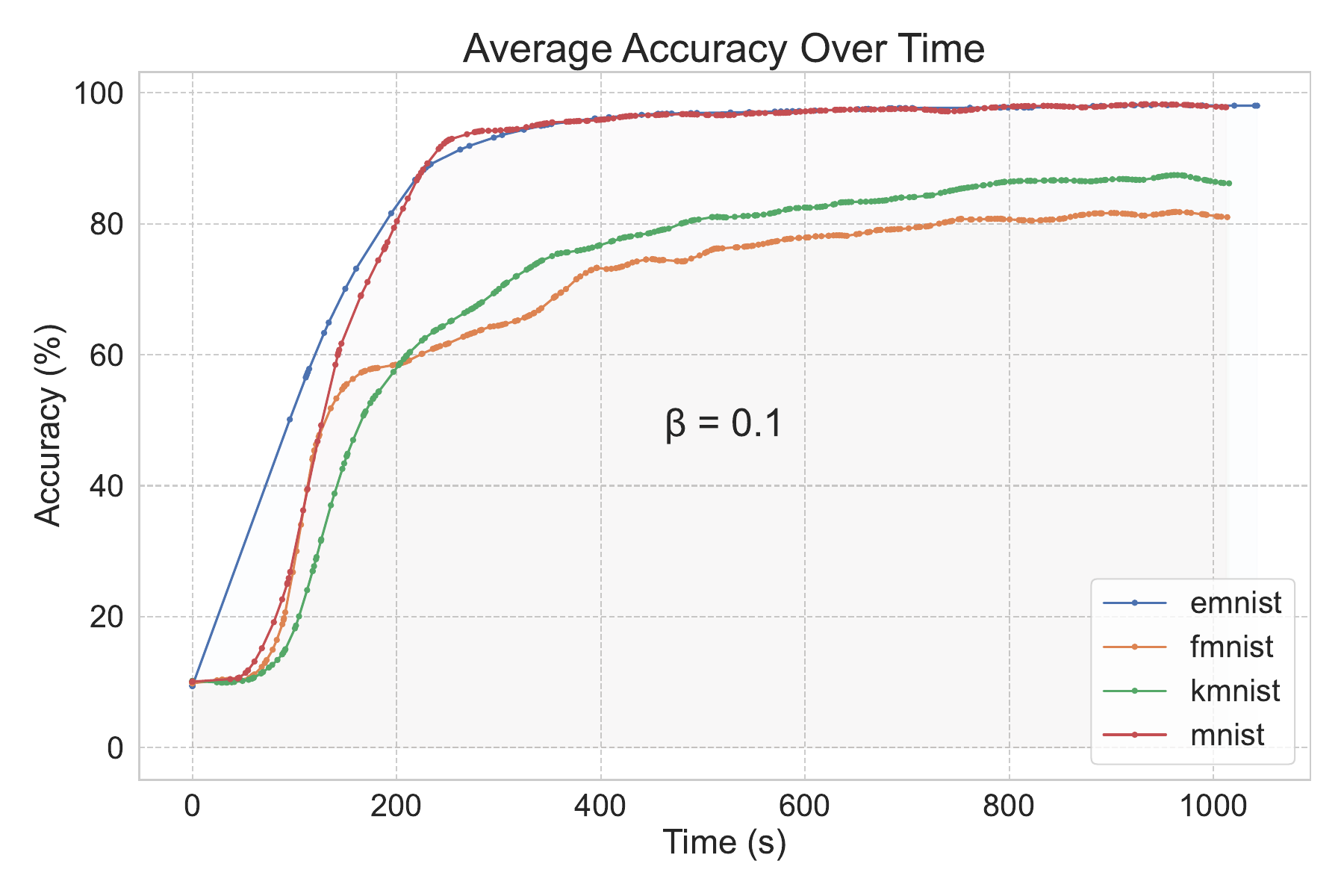}}\\[0.2cm] % spazio verticale ridotto
  % Seconda riga: tre immagini (bar plot)
  \subfloat{\includegraphics[width=0.32\textwidth]{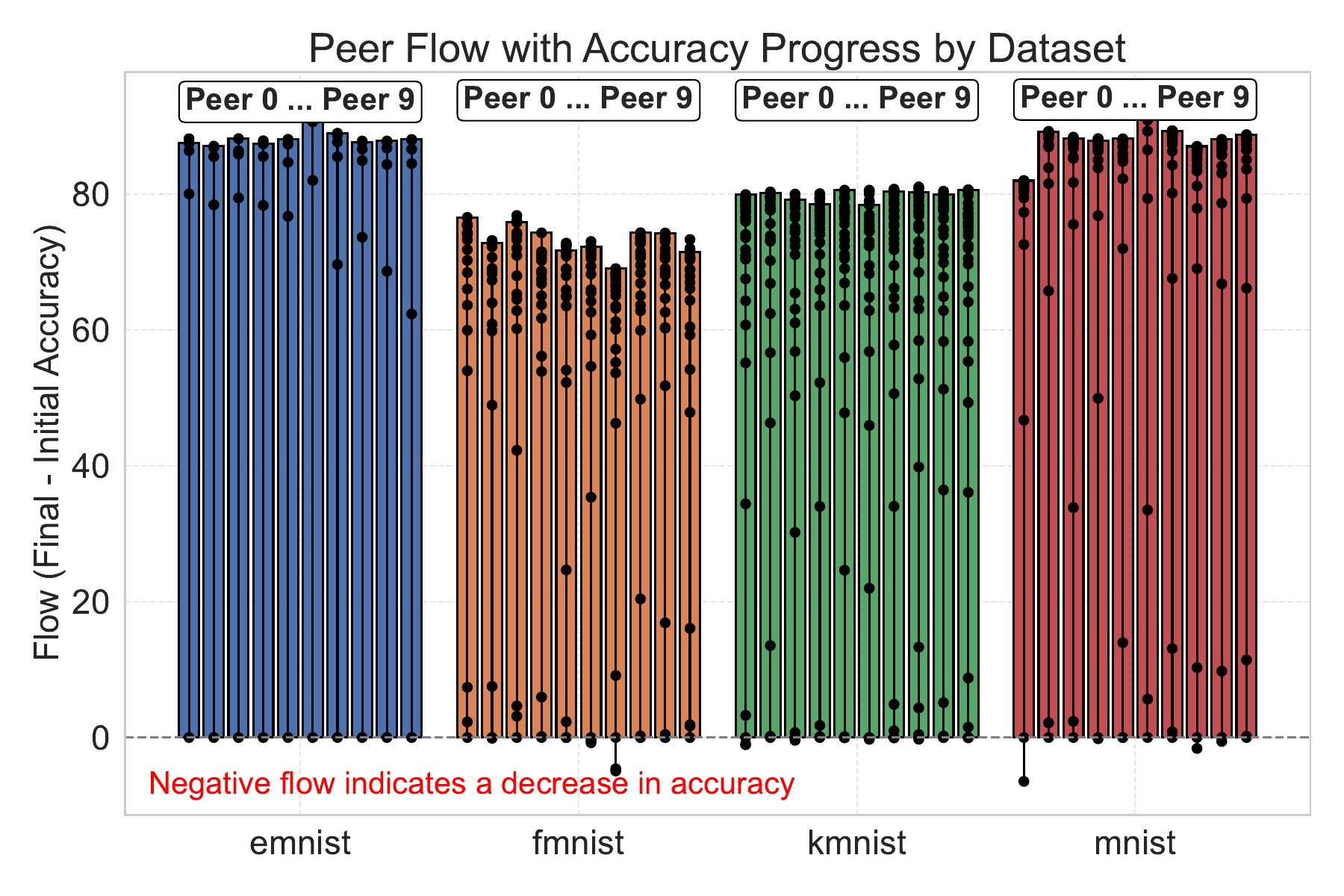}}\hfill
  \subfloat{\includegraphics[width=0.32\textwidth]{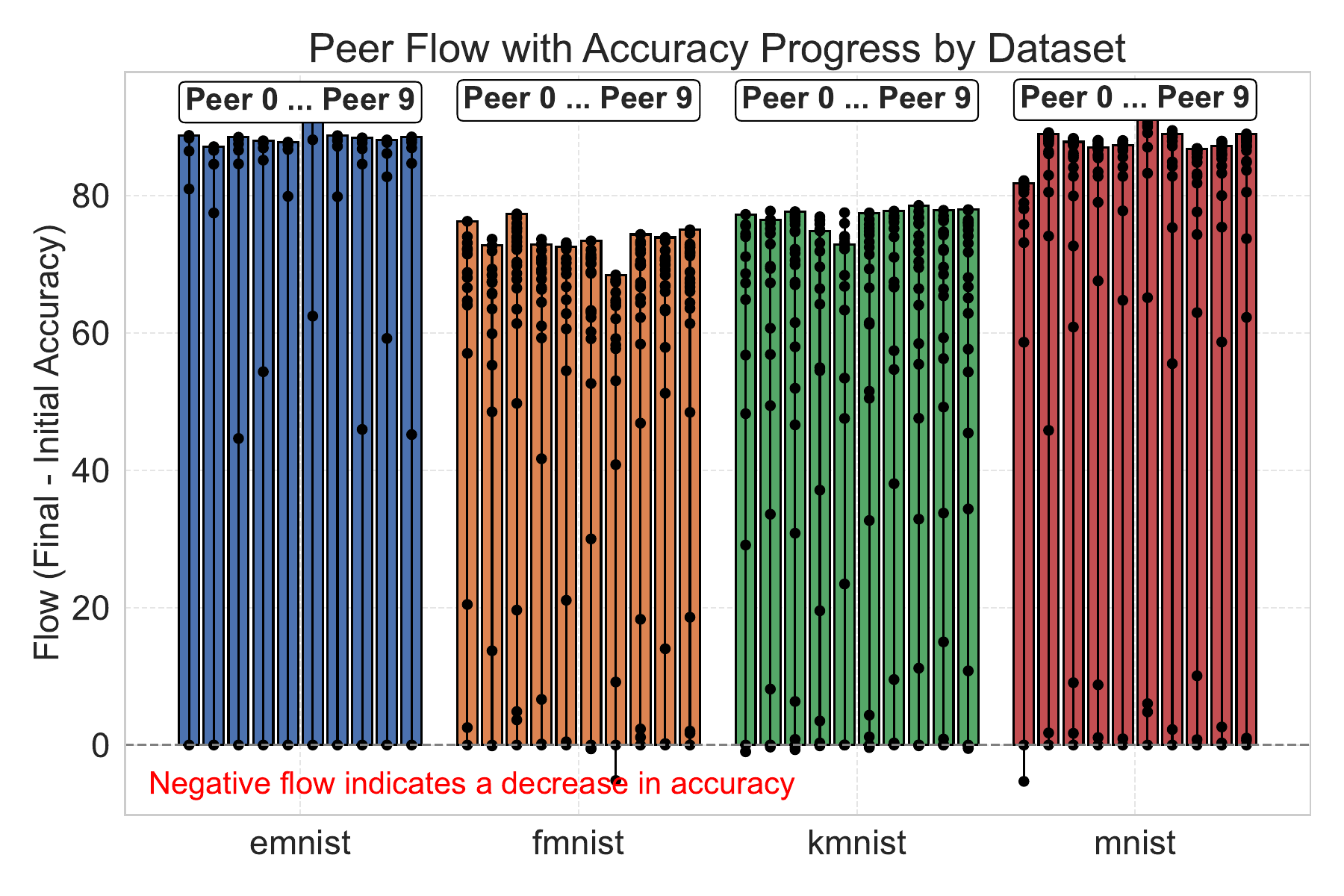}}\hfill
  \subfloat{\includegraphics[width=0.32\textwidth]{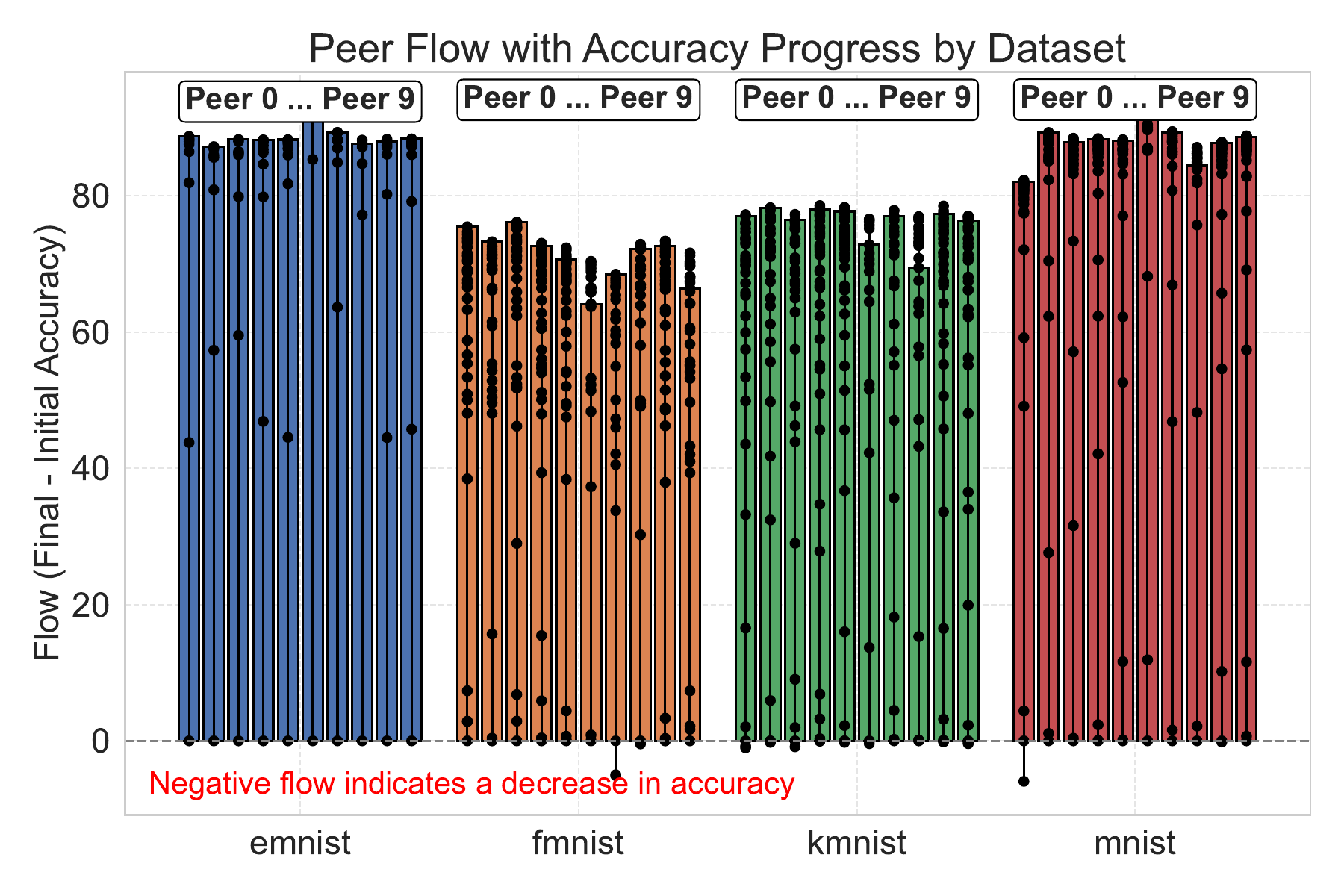}}
  \caption{In the three line charts above, the average accuracies recorded on all grayscale datasets (MNIST and variants) are reported; from left to right, the results correspond to $\beta$=1.0, $\beta$=0.5, and $\beta$=0.1 (the latter under conditions of high heterogeneity among peers). In the lower section, the respective bar plots are shown (each bar plot corresponds to the line chart above it), representing the final and intermediate values for each individual peer.}
  \label{fig:combined}
\end{figure*}

\begin{figure*}[htb]
  \centering
  % Prima riga: tre immagini (loss)
  \subfloat{\includegraphics[width=0.32\textwidth]{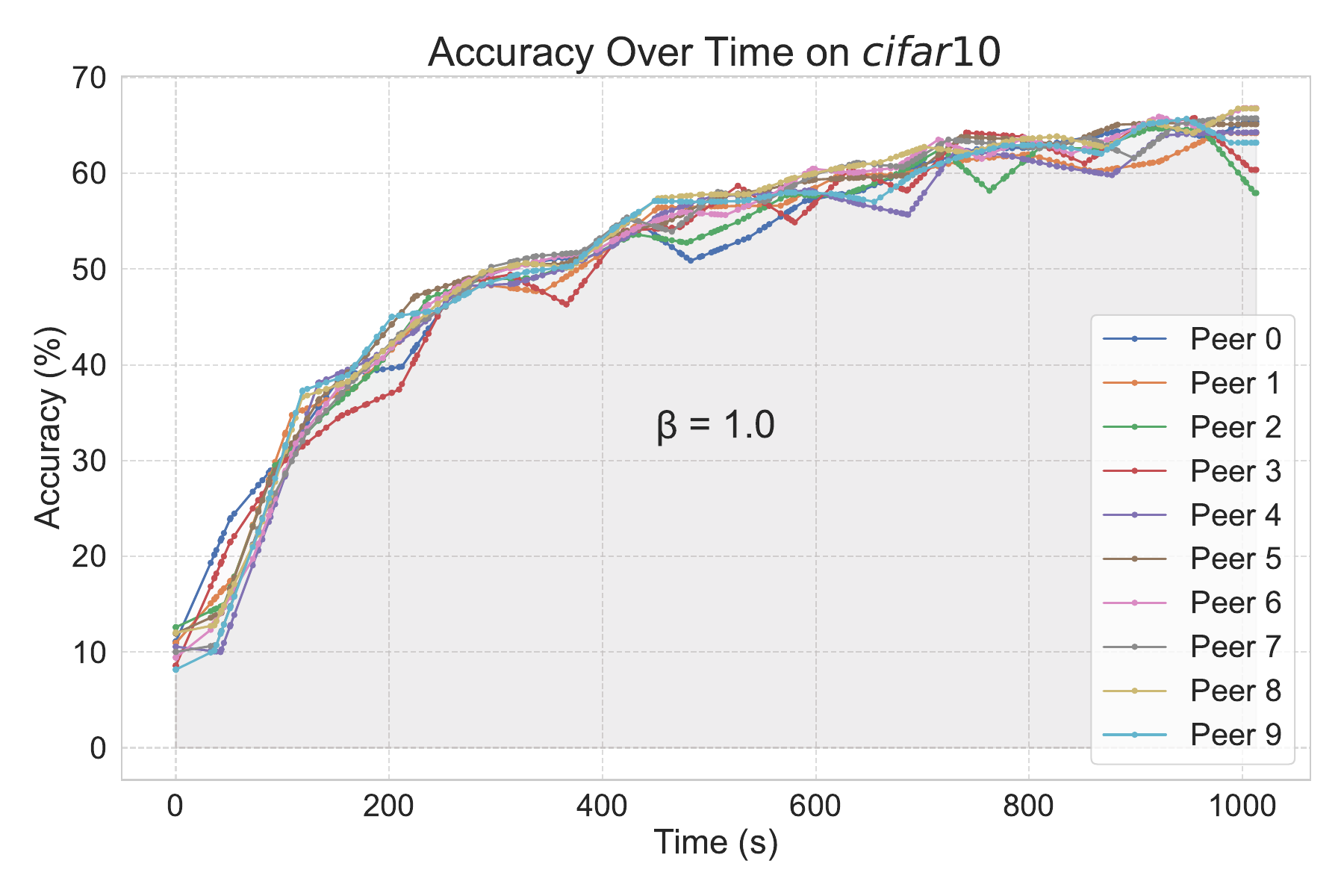}}\hfill
  \subfloat{\includegraphics[width=0.32\textwidth]{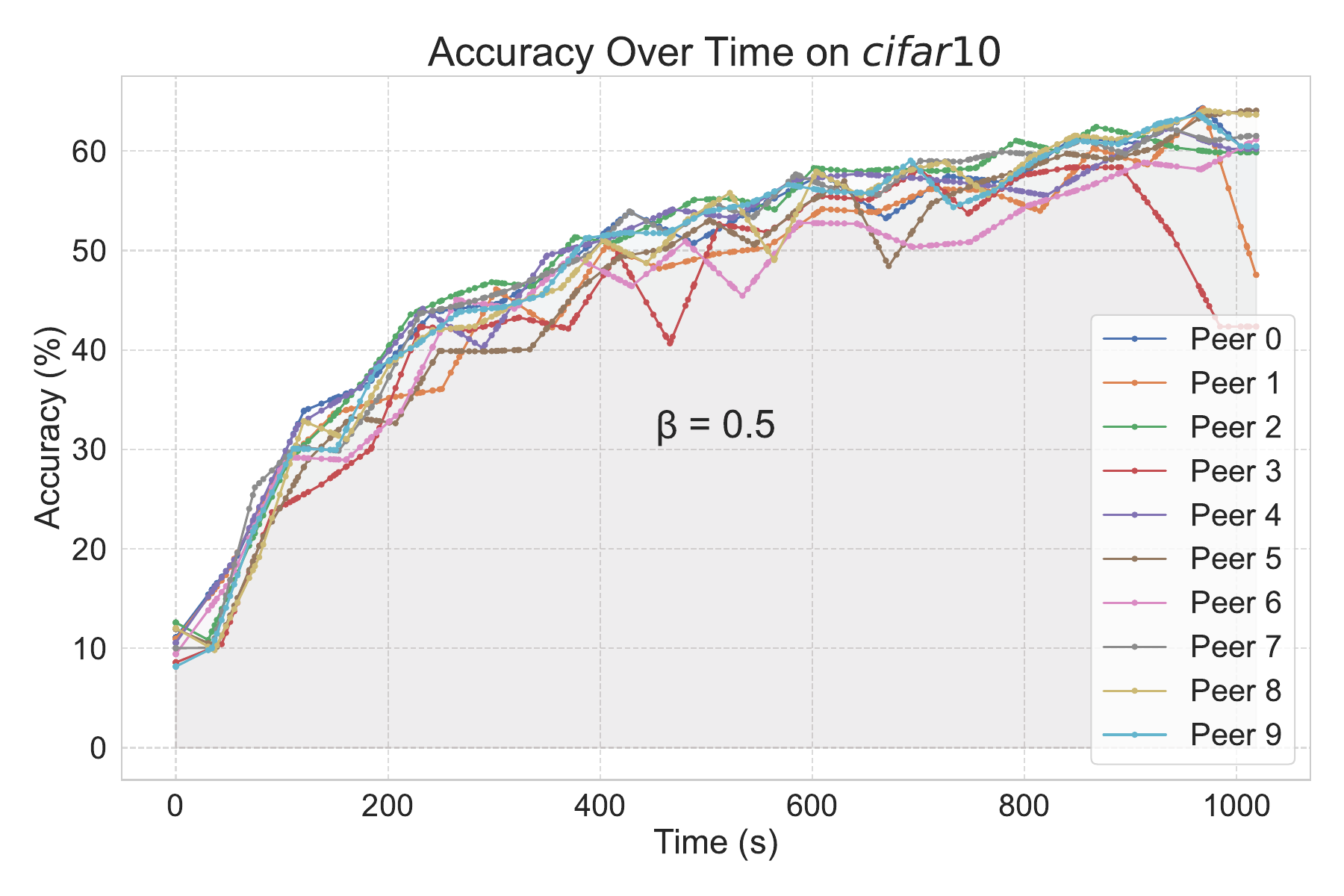}}\hfill
  \subfloat{\includegraphics[width=0.32\textwidth]{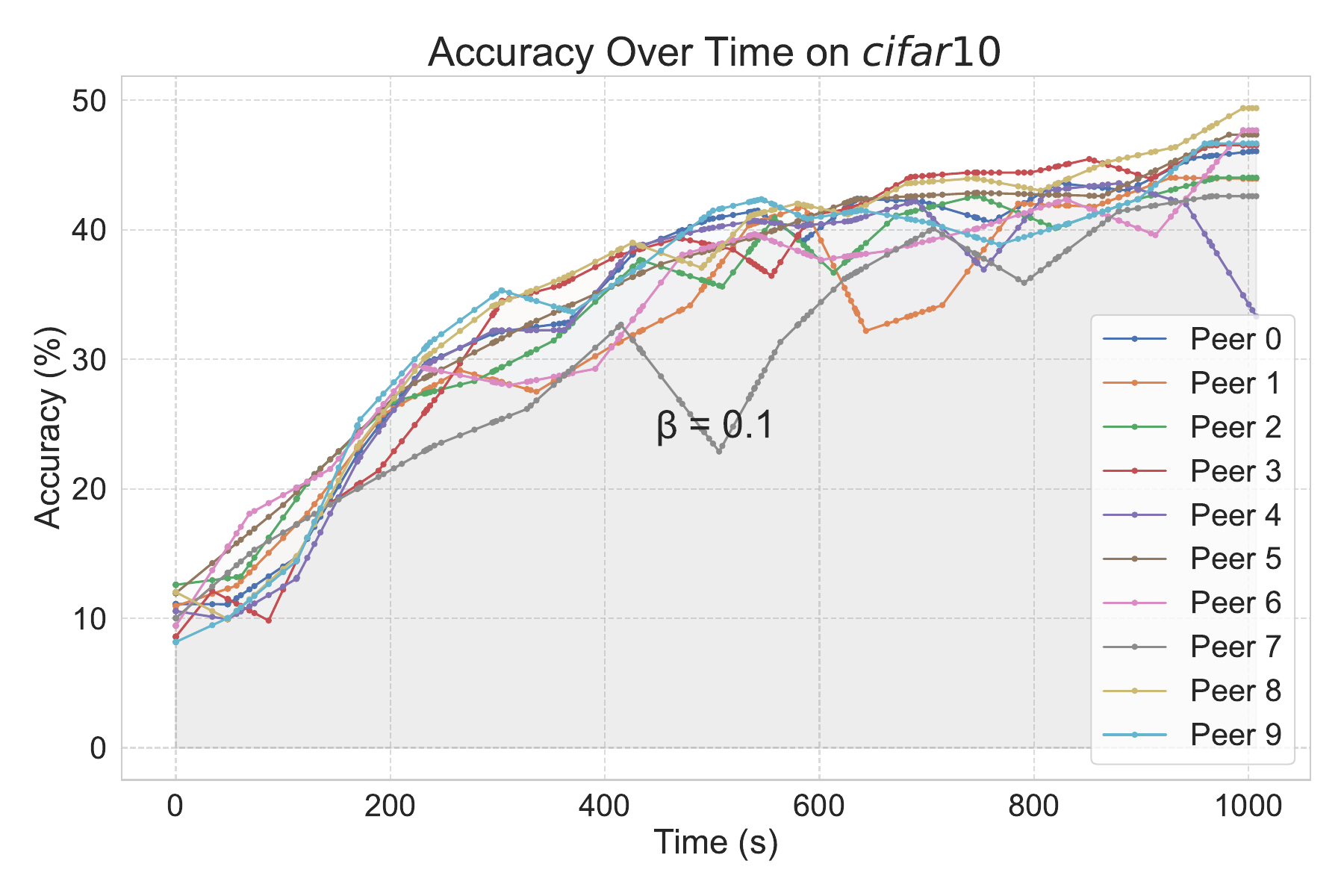}}\\[0.2cm] % spazio verticale ridotto
  \caption{Accuracy trend for each peer on the \textit{CIFAR-10} dataset. As in the previous line charts, performances with three different $\beta$ values (1.0, 0.5, 0.1) are highlighted to demonstrate the robustness of FedBGS when using more complex RGB datasets.}
  \label{fig:combined-cifar}
\end{figure*}

In the three accuracy graphs reported in the Figure \ref{fig:combined}, it is shown that as the $\beta$ parameter varies (the smaller the value of $\beta$, the more heterogeneous the distributions among peers are), the overall training quality remains almost unchanged. The same is not the case, for instance, with FedAvg, which suffers a significant drop in performance as $\beta$ decreases. What we want to demonstrate with these results is that training performance comparable to state-of-the-art methods can be achieved with a fully decentralized automated system, without trade-offs in terms of scalability, costs, or security. The use of a public Ethereum-based blockchain is one of the key elements in this decentralization, a feature that could not be achieved with a private blockchain since it is still managed by a single entity.
The bar plots more clearly show the alignment of the various peers in terms of accuracy; the accuracy achieved, in itself, is not important in this experiment, as it is strictly influenced by the available hardware resources and the use case, namely the network used with its respective hyperparameters and the privacy budget required to ensure a sufficient level of system security, a budget that is closely tied to the dataset and the amount of data available. The parameter by which we will evaluate the effectiveness of the system will be the accuracy growth delta, in addition to a direct theoretical comparison with state-of-the-art systems on the key aspects of FedBGS. Another important aspect highlighted by the bar plots concerns the convergence speed of individual clients' alignment: the black markers indicated in each plot represent the accuracy history for each gossip iteration, and their sparsity in cases with lower $\beta$ values signifies a slower convergence speed, as well as a generally lower accuracy. Resolving the trade-offs between privacy, scalability during clustering, and accuracy will be crucial in the two design phases of FedBGS, depending on the size and type of datasets involved.
Figure \ref{fig:combined-cifar}, on the other hand, shows the accuracy values recorded on each individual peer for the cifar-10 dataset. Despite the classification challenges posed by an RGB dataset compared to the MNIST variants, the model proves to be very robust during convergence, even in the presence of non-IID data.

\begin{table}[htbp]
\centering
\caption{FedBGS Blockchain Usage with Segmented Gossip Learning}
\resizebox{1.0\columnwidth}{!}{%
\begin{tabular}{|c|c|c|c|}
\hline
\textbf{Operation} & \multicolumn{3}{|c|}{\textbf{Gas Usage (Gas Unit)$^{\mathrm{a}}$}} \\
\cline{2-4} 
\textbf{Type} & \textbf{\textit{Interaction Cost}} & \textbf{\textit{Times}} & \textbf{\textit{SC Type}} \\
\hline
SC Deployment           & 1.418.084   & Once            & \#1  \\
\hline
SC\#2 Deployment        & 1.566.634   & Once            & \#2  \\
\hline
Registration            & 100.340    & for Each Client & \#1  \\
\hline
Reset Balance (Optional)& 257.032    & for Each Client & \#2  \\
\hline
Token Penalization      & 77.102     & Variable        & \#2  \\
\hline
Save Hash               & 50.527     & for Each Client & \#2  \\
\hline
Save Cluster Centers    & 257.000    & Once per Update & \#1  \\
\hline
Assign Segment to Peer  & 120.450    & for Each Client & \#1  \\
\hline
Retrieve Segment Boundaries & 35.210   & Each Training Round & \#1  \\
\hline
Validate Segment Update & 65.800     & Variable (Each Update) & \#2  \\
\hline
\multicolumn{4}{l}{$^{\mathrm{a}}$Note that the gas usage also depends on the smart contract's optimization.}
\end{tabular}%
}
\label{tab::Gas Usage}
\end{table}

In the table \ref{tab::Gas Usage}, we show the Ethereum gas used by each function involved in both phases of FedBGS. The functions are split into two types of smart contracts as illustrated in the figure \ref{fig:combined}: \#1 is dedicated to registration and clustering functions, while \#2 is used for the gossiping and penalty phase. While our native choice for FedBGS was Ethereum, as it represents a standard providing high levels of security and functionalities, our gossiping system is designed to support any blockchain that offers advanced capabilities for dApps, such as Cardano, Solana, or PolkaDot. Hence, any blockchain of this type can be a viable choice for implementing FedBGS, with the selection depending on the specific requirements of one’s system.

\section{Conclusions}
To conclude, it can be stated that FedBGS achieved excellent results in terms of convergence by proposing a robust decentralized learning method, while addressing several open challenges in federated learning including full decentralization and the management of heterogeneously distributed data among participants. Furthermore, full decentralization is of fundamental importance given users’ reluctance to trust a centralized structure, and many state-of-the-art methods that propose “serverless" architectures present significant trade-offs in terms of scalability (a problem encountered in many blockchain-based approaches). Future developments of FedBGS will focus on limiting the impact of local differential privacy which, in the presence of few data, significantly degrades overall performance despite scheduling and on methods to adapt FedBGS to constrained architectures, a challenge that proves even more demanding in the context of gossip learning due to the absence of a server.

\section*{Acknowledgment}
This research has been supported by the project “A catalyst for EuropeaN ClOUd Services in the era of data spaces, high-performance and edge computing(NOUS)”, Grant Agreement Number 101135927.

\bibliographystyle{IEEEtran}
\bibliography{bib/federated}

\begin{thebibliography}{00}
\bibitem{b1} G. Eason, B. Noble, and I. N. Sneddon, ``On certain integrals of Lipschitz-Hankel type involving products of Bessel functions,'' Phil. Trans. Roy. Soc. London, vol. A247, pp. 529--551, April 1955.
\bibitem{b2} J. Clerk Maxwell, A Treatise on Electricity and Magnetism, 3rd ed., vol. 2. Oxford: Clarendon, 1892, pp.68--73.
\bibitem{b3} I. S. Jacobs and C. P. Bean, ``Fine particles, thin films and exchange anisotropy,'' in Magnetism, vol. III, G. T. Rado and H. Suhl, Eds. New York: Academic, 1963, pp. 271--350.
\bibitem{b4} K. Elissa, ``Title of paper if known,'' unpublished.
\bibitem{b5} R. Nicole, ``Title of paper with only first word capitalized,'' J. Name Stand. Abbrev., in press.
\bibitem{b6} Y. Yorozu, M. Hirano, K. Oka, and Y. Tagawa, ``Electron spectroscopy studies on magneto-optical media and plastic substrate interface,'' IEEE Transl. J. Magn. Japan, vol. 2, pp. 740--741, August 1987 [Digests 9th Annual Conf. Magnetics Japan, p. 301, 1982].
\bibitem{b7} M. Young, The Technical Writer's Handbook. Mill Valley, CA: University Science, 1989.
\end{thebibliography}

\end{document}